# Interference Mitigation Using Dynamic Frequency Re-use for Dense Femtocell Network Architectures

Mostafa Zaman Chowdhury, Yeong Min Jang, and Zygmunt J. Haas*
Department of Electronics Engineering, Kookmin University, Seoul, Korea
*Department of Electrical and Computer Engineering, Cornell University, USA
E-mail: {mzceee, yjang}@kookmin.ac.kr

*Abstract*— The next generation network aims to efficiently deploy low cost and low power cellular base station in the subscriber's home environment. For the femtocell deployment, frequency allocation among femtocells and macrocell is big concern to mitigate the interference, and to ensure the best use of the expensive spectrum. There are many sources of interference in integrated femtocell/macrocell networks. Lagging in proper management of interference reduces the system capacity, increases the outage probability, and finally users feel bad quality of experience (QoE). The cost effective interference management technique depends on the size of femtocells deployment. In this paper, firstly we present deployable various possible femtocell network scenarios. We propose the dynamic frequency re-use scheme to mitigate interference for femtocell deployment. For highly dense femtocells, we propose the functionalities of self organizing network (SON) based femtocell network architecture. The outage probability of a femtocell user is analyzed in details. The performances of the proposed schemes for various femtocell deployments are performed using numerical analysis.

## I. INTRODUCTION

The low cost and low power femto access point (FAP) can provide growing demand of high bandwidth and better quality of service (QoS) for the indoor users. Network architecture, handover control, and interference management are the key issues for the cost effective integrated femtocell/macrocell network deployment [1]. The network architecture and the resource management mostly depend on the size of existing femtocell deployment, existing network infrastructure, and future extension plan. The interference of femtocell networks cannot be fully minimized. However, it is possible to mitigate the interference. To obtain the best uses of valuable resources (frequency and bandwidth) and user's QoE level, proper radio resource management (RRM) techniques for interference mitigation are needed. A proper interference management increases the system capacity, and reduces the outage probability. The frequency allocation and the power control are the key factors to mitigate interference. The deployable femtocell zones are classified according to the density of femtocells, and co-existence of femtocells and macrocell to mitigate different sources of interferences and to ensure the best utilization of the spectrum.

The interference mitigation using same technique [2], [3]-[5] of frequency allocation for all the dense and discrete femtocell network deployment is not cost effective and cannot ensure the best use of the expensive frequency spectrum. However, the dynamic frequency re-use scheme proposed in this paper increases the frequency spectrum utilization and provides excellent performance. In this scheme, the frequency allocation for the femtocells within a macrocell sector and overlaid macrocell sector use completely separate frequency bands to mitigate femtocell-macrocell interference. The total frequency band allocation for the femtocells within a sector is the sum of the frequency bands of the non-overlapping macrocell sectors. The frequency and power among the neighbor femtocells are automatically adjusted to mitigate the inter-femtocell interference. Self organizing network (SON) based dense FAPs performs self configuration, self optimization, and self healing [6] for the frequency auto configuration and power optimization to mitigate interference effect from the neighbor femtocells. Dynamic re-use frequency scheme is used for the medium and highly dense femtocell network deployment to increase the spectral efficiency and interference avoidance.

The rest of this paper is organized as follows. Section II shows the deployable various femtocell network scenarios and interference scenarios. Interference mitigation using dynamic frequency re-use scheme is proposed in Section III. Section IV provides the outage probability analysis. The numerical analyses for our proposed scheme are shown in Section V. Finally, in Section VI, conclusions are drawn.

## II. INTERFERENCE SCENARIOS FOR FEMTOCELL NETWORKS

Different sources of interferences are found in the femtocell network deployment due to the co-existence of macrocell and femtocells. Thus, the amount of interference depends on the network architecture, location of femtocells, and density of femtocells. Based on these factors following scenarios for the femtocell networks deployment can be explained.

- *Scenario A* **(Single femtocell without overlaid macrocell):** In this case, there is no interference effect from other cells. Normally in very remote area, these types of connections are found.
- *Scenario B* **(Single stand-alone femtocells overlaid by macrocell):** Only discrete femtocells are overlaid by macrocell. In this scenario, there is no femtocell to femtocell interference.

- *Scenario C* (**Multi-femtocells overlaid by macrocell**): Discrete femtocells as well as very few interfering overlapping femtocells are overlaid by macrocell. In this situation, the receivers of macro BS, macro UE, FAP, and femto UE are affected.
- *Scenario D* (**Dense femtocells overlaid by macrocell**): This is the worst case of interference for ultimate goal of femtocell deployment. Lots of femtocells are deployed in a small area. In this case, the receivers of macro BS, macro UE, FAP, and femto UE are affected.

Type of aggressors and victims depend on different situation; position of FAP, macro UE, femto UE, and macro BS. The aggressors are macrocell downlink, macrocell uplink, femtocell downlink, and femtocell uplink [7]. The victims are macro UE receiver, macro BS receiver, femto UE receiver, and FAP receiver.

*Macrocell downlink:* The entire femtocells UEs receive interference from this macrocell downlink. Whenever the location of the femtocell is close to the macro BS and UE is located at the edge of the femtocell, the transmitted power from the macro BS causes interference for the femto UE receiver. All the scenarios *B, C, and D* suffer interference problem for the Macrocell downlink.

*Macrocell uplink:* FAP receiver suffers from this aggressor. Whenever the macro UE is close to the femtocell or inside the femtocell coverage area, the transmitted uplink signal from the macro UE to macro BS causes the interference for the FAP receiver. All the scenarios *B, C,* and *D* suffer this interference problem.

*Femtocell downlink:* The femtocell downlink causes the interference for the macro UE receiver and nearby femto UE receiver. Whenever a macro user is very near, or inside a femtocell coverage area, the macro UE feels interference for this femtocell downlink (scenarios *B, C,* and *D*). Also, if two or more femtocells are very close, then the femto UE of one femtocell receives interference from the neighbor femtocell downlink (scenarios *C* and *D*).

*Femtocell uplink:* Macrocell receiver and nearby FAP receiver suffer from this aggressor. Whenever the femtocell is close to the macrocell, the transmitted uplink signal from the femto UE to FAP causes the interference for the macrocell receiver (scenarios *B, C,* and *D*). Also, if two femtocells are close to each other, the femtocell uplink causes interference for the neighbor FAP receiver (scenarios *C* and *D*).

III. PROPOSED SCHEMES TO MITIGATE INTERFERENCE FOR FEMTOCELL NETWORKS

The spectrum management for femtocells and macrocell, and the power management are the main controlling factors for the interference mitigation. The investigation of co-channel interference mitigation techniques (such as, interference cancellation, interference randomization, and interference avoidance) has become a key focus area in achieving dense spectrum re-use in next generation cellular systems [8]. Different proposed techniques for interference mitigation are explained in this section.

A. *Frequency Configuration*

In cellular wireless communication system, the frequency is very expensive. The total cellular frequency band (say, $f_1$ to $f_2$) allocation for different femtocell deployments should be different for the efficient uses of frequencies. This section proposes different frequency allocation for different scenarios of femtocell deployment.

Let the total cellular frequency band, allocated frequency band for macrocell, and allocated frequency band for femtocells be denoted as $B_T$, $B_m$, and $B_f$ respectively. The macrocell and its subsectors may use *p* numbers of sub-bands. Whereas, all the femtocells within a macrocell may use *q* numbers of sub-bands. The index *x* and *y* indicates upper and lower frequency of each band or sub-band respectively.

$$B_T = f_2 - f_1 \quad (1)$$

$$B_m = B_m \in B_T = \sum_{b=1}^{p} B_{mb} = \sum_{b=1}^{p} (f_{mxb} - f_{myb}) \quad (2)$$

$$B_f = B_f \in B_T = \sum_{b=1}^{q} B_{fb} = \sum_{b=1}^{q} (f_{fxb} - f_{fyb}) \quad (3)$$

- *Dedicated Frequency Band*

In this scheme, femtocell and macrocell uses totally separate frequency band. Fig. 1 shows that all the femtocells use the frequency band from $f_1$ to $f_3$, whereas total frequency band allocation for the overlaid macrocell is $f_3$ to $f_2$. All the femtocells will use the same frequency. This scheme will not support dense femtocell deployment (scenarios *D*) and will not efficient for scenarios *A*. The same frequency among the dense neighbor femtocells will cause more interference. The scenarios *A* will cause wastage of frequency. This scheme can be used for the scenarios *B* only.

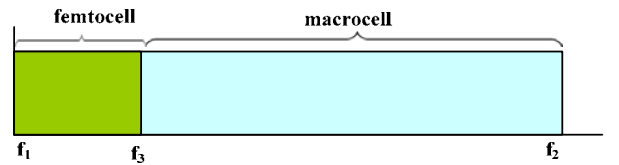

**Fig.1.** Different frequency band is allocated for femtocells and macrocell

- *Same Frequency Band*

Same frequency band is allocated for both the femtocells and macrocell in this scheme. Fig. 2 shows the frequency allocation for femtocells and macrocell.

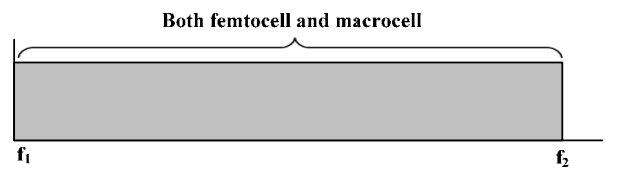

**Fig.2.** Same frequency is allocated for both the femtocells and macrocell

This scheme is very much efficient for the scenarios *A* to make the best utilization of frequencies. For the other scenarios, this scheme causes interference. However, it can be used for scenario *B* for very less number of discrete femtocells overlaid by macrocell.

- *Partial Frequency Band*

In this scheme, macrocell uses total allocated frequency band, and a part of total frequency band is used by the femtocells. Fig. 3 shows that all the femtocells use frequency band $f_1$ to $f_4$, whereas total frequency allocation for the macrocell is $f_1$ to $f_2$. All the femtocells will use the same frequency. This scheme will not support dense femtocell deployment (scenario *D*). This scheme will also make huge interference for scenarios *C*. However this scheme causes not much interference scenarios *B*. The same frequency among the dense neighbor femtocells will cause more interference. This scheme is more efficient for the scenario *A*.

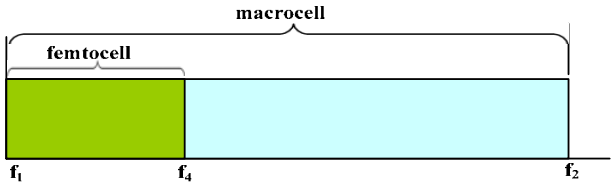

**Fig.3.** Partial frequency is allocated for femtocells and macrocell

- Dynamic re-use frequency

In the large scale deployment of femtocell networks, frequency re-use is ultimate choice for the frequency allocation to make the best uses of frequencies. This sub-section provides the proposed dynamic frequency re-use scheme.

Total cellular frequency is divided into three equal parts. Each of the three sectors of the macrocell uses any of the three different frequency band $B_{m1}$ or $B_{m2}$ or $B_{m3}$. One sector uses one part and the femtocells within that sector can use the remaining two parts of the frequency band. However, the frequency of each femtocell may contain two separate bands. One band is using for the center frequency and other band for the edge frequency. All the femtocells use the same frequency in the center of the cell. The edge of each neighbor femtocells uses different frequency band to avoid the interference. Fig. 4 shows the frequency allocation for the dynamic frequency re-use scheme. For this scheme

$$\left.\begin{aligned} B_{m1} = B_{m2} = B_{m3} &= \frac{B_m}{3} \\ B_{f1} = B_{m2} + a.B_x + b.B_y &+ c.B_z \end{aligned}\right\} \quad (4)$$

where, $a$, $b$, and $c$ are the binary value (0 or 1) and the value of $(a + b + c)$ is also a binary value. The value of $a$, $b$, and $c$ depends on the neighbor femtocell's edge frequency band. $B_{m1}$, $B_{m2}$, and $B_{m3}$ refers the frequency band allocation for sector 1, 2, and 3 of macrocell respectively. $B_{f1}$ refers the frequency band allocation for each of the femtocell in sector 1 of the macrocell.

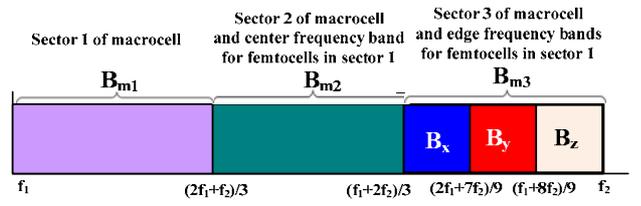

(a) Division of frequency spectrum for dynamic re-use frequency

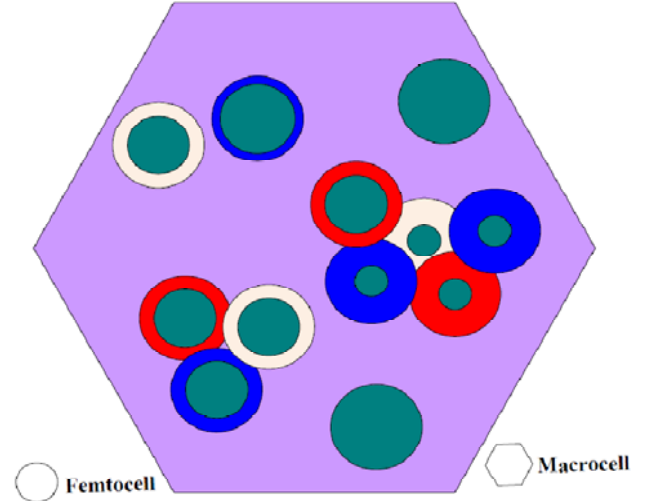

(b) Frequency allocations for dynamic re-use frequency scheme

**Fig. 4.** Frequency allocations among macrocell and femtocells for dynamic re-use frequency scheme

The radius of the inside circle can be varied according to the number of neighbor femtocells and their distance. Scenario *C* and *D* is very much effective for this scheme. The transmitted power can be automatically adjusted and edge frequencies can be automatically configured using the SON based femtocell networks [9]. The macrocell with *N* numbers of sectors can follow the same procedure of frequency allocation.

*B.  Proposed functionalities for SON based femtocell networks to support dynamic frequency re-use scheme*

The main functionalities of SON for integrated femtocell/macrocell networks are: self-configuration, self-optimization, and self-healing. The self-configuration includes intelligent frequency allocation among neighbor FAPs; self-optimization includes optimization of transmission power among neighbor FAPs, neighbor cell list, coverage, and mobility robustness; and self-healing includes automatic detection and solution of most of the failures. The sniffing function is required to integrate the femtocell into a macrocellular network so that FAP can scan the air interface for available frequencies and other network resources. The integrated femtocell/macrocell networks should have SON capabilities so that neighbor FAPs can communicate with each other to configure resources, transmission power, and frequency. Therefore, further enhancements of SON will be an essential element for the future femtocell deployment. We briefly explain three scenarios among many scenarios we expected.

- **Scenario *1* (a frequency configuration and power optimization):** If large number of FAPs are deployed in an indoor building or femto zone area, signals from different FAPs will interfere with each other. These FAPs coordinate with each other to configure frequency and to optimize transmission power.

**Scenario *2* (cell size adjustment ):** If an UE is connected with a FAP, designated as a master FAP, and it receives interference signal from other FAPs, then the master FAP requests those FAPs to configure their transmission power, so interference can be reduced. Also cell size is re-adjusted to reduce the interference effect.

**Scenario *3* (Frequency allocation for a newly installed FAP in a macrocell coverage area):** Whenever a FAP is newly installed in a macrocell coverage area, FAP and macrocellular base station can communicate each other to configure frequency for FAP if needed.

## IV. OUTAGE PROBABILITY ANALYSIS

Poor interference management system causes reduction of system capacity for both the macrocell and femtocell networks. Interference causes the higher outage probability. Finally, the more interference causes reduction of QoE level for the users. A good interference management system can increase the femtocell capacity as well macrocell capacity. The capacity of a wireless channel decreases with decreasing the value of signal-to-interference ratio (SIR). The received SIR of a femtocell user in a femtocell coverage area for macrocell/femtocell overlaying networks can be expressed as

$$SIR = \frac{S_o}{I_m + \sum_{i=1}^{K} I_{n(i)}} \qquad (5)$$

where $S_o$ is the received signal from the expected reference femtocell, $I_m$ is the received interference signal from the reference macrocell, and $I_{n(i)}$ is the received interference signal from the *i-th* femtocell among *K* numbers of neighbor femtocells. The index *0, i,* and *m* refers the reference femtocell, *i-th* neighbor femtocell, and macrocell respectively.

The outage probability can be presented as

$$P_{out} = P_r(SIR < \gamma) \qquad (6)$$

where *γ* is a threshold value of SIR below which a call cannot continue.

Alternatively, (6) can be written as

$$P_{out} = P_r\left(\frac{S_o}{I_m + \sum_{i=1}^{K} I_{n(i)}} < \gamma\right)$$

$$= P_r\left\{\sum_{i=1}^{K} I_{n(i)} > \left(\frac{S_o}{\gamma} - I_m\right)\right\}$$

$$= P_r\left\{I_f > \left(\frac{S_o}{\gamma} - I_m\right)\right\} \qquad (7)$$

where $I_f$ and $I_m$ are the total received interference from the co-channel neighbor femtocells and macrocells respectively. The probability density function (PDF) of, $I_f$ can be assumed as a Gaussian distributed [10].

Equation (7) shows that he received interference signal from interfering macrocell and neighbor femtocells influence the outage probability of a call. The mitigation techniques of these interferences are varied according to availability of macrocell coverage and dense of femtocells. For a remote area where macrocell coverage is very poor or not available; also if the macrocell and femtocells are deployed using separate frequency band then $I_m$ can be assumed to be zero. In femtocell network deployment, a lot of neighbor femtocells are exist. Thus, only proper interference management can increase signal-to-interference ratio and can reduce the outage probability for femtocell network deployment. Hence, the major concern is about the mitigation of interference from the neighbor femtocells.

Interference causes the outage probability of a call. More interference makes the channel condition bad. Thus, the outage probability increases if there is no effective interference management system. In cellular environment, the received power $P_R$ from a transmitter of transmitted power $P_T$ can be written as

$$P_R = P_T P_0 \, d^{-\eta} \xi Z \qquad (8)$$

where $P_0$ is the function of carrier frequency, antenna height, antenna gain. *d* is the distance from transmitter to receiver, *ξ* is the slow fading variable, and *Z* is the fast fading variable. The distribution of *ξ* and *Z* for Rayleigh fading are normally exponential distribution [11]. *η* is the path loss exponent.

When the femtocell user receive signal from its won FAP that is situated at the same indoor environment, the slow fading can be avoided for this condition. Thus, from (8), the received signal by the interested femtocell user from its own femtocell $S_o$, interference $I_{n(i)}$, and $I_m$ from the *i-th* neighbor femtocell and reference macrocell respectively can be expressed as

$$S_o = P_{Rf(0)} = P_{Tf(0)} P_{0f} d_0^{-\eta_1} Z_0 = \bar{S} \, Z_0 \qquad (9)$$

$$I_{n(i)} = P_{Rf(i)} = P_{Tf(i)} P_{0f} d_i^{-\eta_2} \, \xi_i Z_i \qquad (10)$$

$$I_m = P_{Rm} = P_{Tm} P_{0m} d_m^{-\eta_3} \xi_m Z_m \qquad (11)$$

where $i = 0$ means the reference femtocell. *m* refers the reference femtocell. $P_{Rm}$ and $P_{Rf(i)}$ are the received power at MS from the reference macrocell and *i-th* neighbor femtocell respectively. $P_{Tm}$ and $P_{Tf(i)}$ represents the transmitted power from the macrocell and *i-th* femtocell respectively. $d_m$ and $d_i$ are the distance from the interested MS to reference macro BS and *i-th* femtocell respectively.

Now $I_f$ and $I_m$ are the total received interference signal from the *K* numbers of neighbor femtocells and reference macrocell respectively. Using (10) and (11), $I_f$ and $I_m$ can be calculated as

$$I_f = \sum_{i=1}^{K} I_{n(i)} \; = \sum_{i=1}^{K} P_{Tf(i)} P_{0f} d_i^{-\eta_2} \, \xi_i Z_i X_i \qquad (12)$$

$$I_m = P_{Tm} P_{0m} d_m^{-\eta_3} \xi_m Z_m Y \qquad (13)$$

where $X_i$ and $Y$ are the binary random variable, which takes value in {0,1}, presents whether *i-th* neighbor femtocell or reference macrocell respectively and experimented femtocell use same frequency ($X_i = 1$; $Y = 1$) or different frequency ($X_i = 0$; $Y = 0$).

Using (7) and (9), $P_{out}$ can be written as

$$P_{out} = P_r \left\{ Z_0 < \frac{\gamma}{S}(I_f + I_m) \right\} \quad (14)$$

The PDF of $Z_0$ is exponentially distributed. Thus, the solution of (14) to calculate the $P_{out}$ can be found as

$$P_{out} = \int_0^{\frac{\gamma}{S}(I_f + I_m)} \exp(-Z_0) \, dZ_0$$

$$= 1 - \exp\left\{-\frac{\gamma}{S}(I_f + I_m)\right\}$$

$$= 1 - exp\left[-\frac{\gamma}{S}\left\{\sum_{i=1}^{K} I_{n(i)} + I_m\right\}\right]$$

$$= 1 - \left\{\prod_{i=1}^{K} e^{\left\{-\frac{\gamma}{S} I_{n(i)} X_i\right\}}\right\} e^{\left\{-\frac{\gamma}{S} I_m Y\right\}} \quad (15)$$

Equations (9) to (13) are used to calculate the $P_{out}$ in (15). In (15), $Y = 1$ if the reference macrocell and the reference femtocell are deployed with the same frequency, otherwise $Y = 0$ (suppose for $B_m \cap B_f = \emptyset$ ). Thus for the same frequency band and partial frequency band schemes, $Y = 1$ and for other schemes shown in this paper, $Y = 0$. The value of $X_i$ is related to the allocated frequency for the reference femtocell and *i-th* neighbor femtocell. For the same frequency band, dedicated frequency band, and partial frequency band schemes where all the $K$ numbers of neighbor femtocells use same frequency like reference femtocell produces $X_i = 1$. For the dynamic frequency re-use scheme, more than 66% neighbor femtocells produce $X_i = 0$. Thus the outage probability for the dynamic re-use scheme is lower than other schemes.

## V. NUMERICAL RESULTS

In this section, we verified the performance of the proposed dynamic re-use frequency scheme. We assume more than 1000 femtocells within a macrocell as a dense femtocell deployment (scenario "D"). Table 2 shows our basic assumption for the numerical analysis. For the simplicity, existing propagation model for the macrocell in [8] and for femtocell in [2] are used. Only reference macrocell is considered for the simulation. One wall between two femtocells is considered to calculate inter femtocell interference. We consider the femtocells within 100 meter range as neighbor femtocells. The positions of the femtocells within the macrocell coverage area are placed randomly and the number of femtocells within the neighbor area is randomly generated according to Poisson distribution. For the dedicated frequency scheme, we assume 33.3% of total cellular frequency band is allocated for femtocells and remaining 66.6% of frequency band is allocated for the macrocell.

TABLE 2. BASIC ASSUMPTIONS FOR THE ANALYSIS

| Parameter | Assumption |
|---|---|
| Radius of macrocell | 1000 m |
| Radius of femtocell | 10 m |
| Distance between the reference macro BS and the experimented FAP | 200 m |
| Carrier frequency | 900 MHz |
| Transmit signal power by macro BS | 1.5 W |
| Maximum transmit signal power by FAP | 10 mW |
| Height of macro BS | 50 m |
| Height of FAP | 2 m |
| Threshold value of SIR ($\gamma$) | 9 dB |

Fig. 5 shows that, proposed dynamic re-use frequency scheme reduces the outage probability within a reasonable range when we consider 1000 random femtocells within the macrocell coverage area. Fig. 6 shows that, outage probability for the proposed dynamic re-use frequency scheme is very small compared to the other schemes even for very high dense femtocell network deployment.

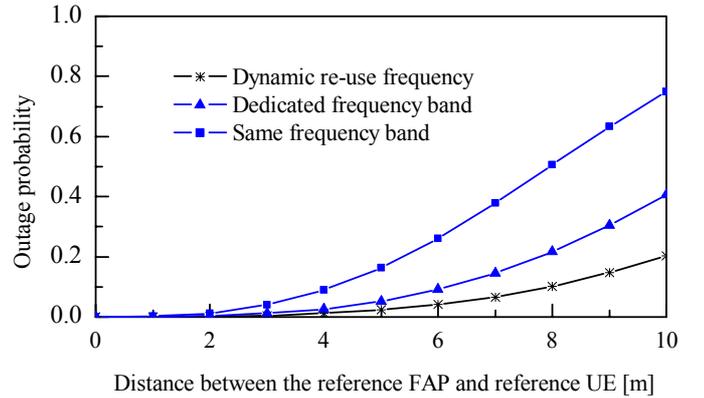

**Fig. 5.** Outage probability comparison when 1000 random femtocells are considered within the macrocell coverage area

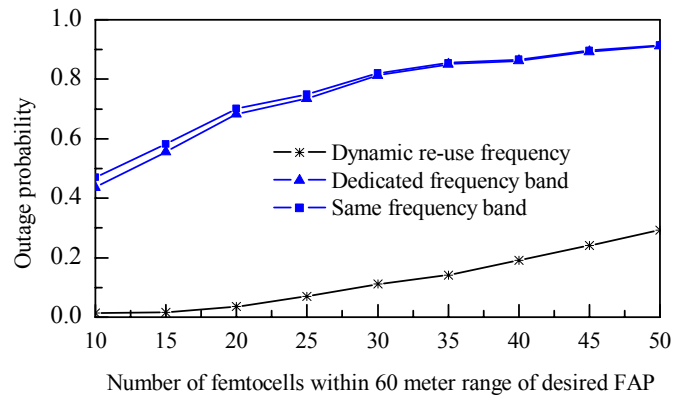

**Fig. 6.** Outage probability for dense femtocell scenario. The distance between the femto UE and the desired FAP is always 5 meter

Actually dynamic frequency re-use scheme provides excellent performance for all of scenario *B, C,* and *D*. However, dynamic frequency re-use scheme is little expensive for the scenario *B* where very small numbers of femtocells are deployed*,* because, dynamic frequency re-use scheme must need SON based network architecture. Hence, proposed dynamic frequency re-use scheme provides outstanding performance in terms of higher throughput, lower outage probability, and utilization of expensive spectrum for the dense femtocells which is the ultimate goal of femtocell network deployment.

## VI. CONCLUSIONS

Both the femtocell and macrocell use the same cellular frequency band. Thousands of femtocells are overlaid by a macrocell. Therefore interference is a key issue for the femtocell network deployment. The effects of poor management of interference are discussed here in details. Different femtocells deployable scenarios are shown in this paper. Based on the existence of the numbers of femtocells within a macrocell and their location in the macrocell, different sources of interference are found. The proposed dynamic frequency re-use allocation scheme is able to mitigate the interference properly with better spectral efficiency and lower outage probability. The proposed SON based femtocell network architecture is very much effective for the future dense femtocell network deployment. The numerical results show that, the proposed scheme is able to mitigate the interference which reduces the outage probability. Thus, our proposed cost effective interference mitigation using dynamic frequency re-use technique is very much promising for the full scale of femtocell network deployment.

The effects of multiple macrocells are not considered in this paper. For the future work multiple macrocells and capacity analysis will be considered. Future work will also consider static frequency re-use scheme for the discrete femtocells.

ACKNOWLEDGEMENT

This research was supported by the MKE (Ministry of Knowledge and Economy), Korea, under the ITRC (Information Technology Research Center) support program supervised by the IITA (Institute of Information Technology Assessment), 2010.